\documentstyle[epsfig,subfigure,12pt]{article}
\catcode`\@=11 % This allows us to modify PLAIN macros.
\let\rel@x=\relax
\newcount\timecount
\newcount\hours \newcount\minutes  \newcount\temp \newcount\pmhours

\hours = \time
\divide\hours by 60
\temp = \hours
\multiply\temp by 60
\minutes = \time
\advance\minutes by -\temp
\def\hour{\the\hours}
\def\minute{\ifnum\minutes<10 0\the\minutes
            \else\the\minutes\fi}
\def\clock{
\ifnum\hours=0 12:\minute\ AM
\else\ifnum\hours<12 \hour:\minute\ AM
       \else\ifnum\hours=12 12:\minute\ PM
            \else\ifnum\hours>12
                 \pmhours=\hours
                 \advance\pmhours by -12
                 \the\pmhours:\minute\ PM
                 \fi
            \fi
         \fi
\fi
}

\def\monthname{\rel@x\ifcase\month 0/\or January\or February\or
   March\or April\or May\or June\or July\or August\or September\or
   October\or November\or December\else\number\month/\fi}

\def\bold#1{\setbox0=\hbox{$#1$}%
     \kern-.025em\copy0\kern-\wd0
     \kern.05em\copy0\kern-\wd0
     \kern-.025em\raise.0433em\box0 }

\def\lsim{\mathrel{\mathpalette\@versim<}}
\def\gsim{\mathrel{\mathpalette\@versim>}}
\def\@versim#1#2{\vcenter{\offinterlineskip
        \ialign{$\m@th#1\hfil##\hfil$\crcr#2\crcr\sim\crcr } }}

\newcount\@tempcntc
\def\@citex[#1]#2{\if@filesw\immediate\write\@auxout{\string\citation{#2}}\fi
  \@tempcnta\z@\@tempcntb\m@ne\def\@citea{}\@cite{\@for\@citeb:=#2\do
    {\@ifundefined
       {b@\@citeb}{\@citeo\@tempcntb\m@ne\@citea\def\@citea{,}{\bf ?}\@warning
       {Citation `\@citeb' on page \thepage \space undefined}}%
    {\setbox\z@\hbox{\global\@tempcntc0\csname b@\@citeb\endcsname\relax}%
     \ifnum\@tempcntc=\z@ \@citeo\@tempcntb\m@ne
       \@citea\def\@citea{,}\hbox{\csname b@\@citeb\endcsname}%
     \else
      \advance\@tempcntb\@ne
      \ifnum\@tempcntb=\@tempcntc
      \else\advance\@tempcntb\m@ne\@citeo
      \@tempcnta\@tempcntc\@tempcntb\@tempcntc\fi\fi}}\@citeo}{#1}}
\def\@citeo{\ifnum\@tempcnta>\@tempcntb\else\@citea\def\@citea{,}%
  \ifnum\@tempcnta=\@tempcntb\the\@tempcnta\else
   {\advance\@tempcnta\@ne\ifnum\@tempcnta=\@tempcntb \else \def\@citea{--}\fi
    \advance\@tempcnta\m@ne\the\@tempcnta\@citea\the\@tempcntb}\fi\fi}
\catcode`\@=12 % at signs are no longer letters

\begin{document}
\vskip 20pt
\def\beq{\begin{equation}}
\def\eeq{\end{equation}}
\def\bea{\begin{eqnarray}}
\def\eea{\end{eqnarray}}
\def\MSbar {\hbox{$\overline{\hbox{MS}}\,$}}
\def\smallMSbar {\hbox{$\overline{\hbox{{\footnotesize MS}}}\,$}}
\def\aBj{\hbox{$a^*_{Bj}$}}
\def\Obs{\hat{O}}
\def\CI{{\cal C}}
\def\naive{na\"{\i}ve}
\def\etal{{\em et al.}}
\newcommand{\mycomm}[1]{\hfill\break{ \tt===$>$ \bf #1}\hfill\break}
\newcommand{\eqref}[1]{eq.~(\ref{#1})}   % for referencing eqs by name
\newcommand{\eqrefA}[1]{(\ref{#1})}   % for referencing eqs by name
\def\newline{\hfill\break}
\def\tfrac#1#2{{#1/#2}} % to avoid fractions in running text
\def\ra{\hbox{$\rightarrow$}\ }

\parskip 0.3cm

\begin{titlepage}
\begin{flushright}
TAUP-2393-96\\
%hep-ph/9611453
\end{flushright}

\begin{centering}
{\large{\bf
Why Pad\'e Approximants reduce the Renormalization-Scale dependence in QFT?
}}\\
\vspace{1.8cm}
{\bf Einan Gardi} \\
\vspace{.15in}
School of Physics and Astronomy
\\ Raymond and Beverly Sackler Faculty of Exact Sciences
\\ Tel-Aviv University, 69978 Tel-Aviv, Israel
\\ e-mail: gardi@post.tau.ac.il
\\
\vspace{1.8cm}
{\bf Abstract} \\

\bigskip
{\small
We prove that in the limit where the $\beta$ function is dominated by
the 1-loop contribution (``large $\beta_0$ limit'') diagonal
Pad\'e Approximants (PA's) of perturbative series
become {\em exactly} renormalization scale (RS) independent.
This symmetry suggest that diagonal PA's are resumming correctly
contributions from higher order diagrams that are responsible for the
renormalization of the coupling-constant.
Non-diagonal PA's are not exactly invariant, but
generally reduce the RS dependence as compared to partial-sums.
In physical cases, higher-order corrections in the $\beta$ function break
the symmetry softly, introducing a small scale and scheme dependence.
We also compare the Pad\'e resummation with the
BLM method. We find that in the large-$N_f$
limit using the BLM scale is identical to resumming the series by a
$x[0/n]$ non-diagonal PA.
} % end of \small
\end{centering}
%\vspace{0.5in}
\vfill
\end{titlepage}

\section{Introduction}

Pad\'e Approximants (PA's) have proven to be useful in many physics
applications, including condensed-matter problems and
quantum field theory \cite{padeworks}.
We denote {\em Pad\'e Approximants} (PA's)
to a perturbative QCD series, describing a generic effective-charge 
$S(x) = x(1\,+\,r_1 x \,+\,r_2 x^2\, +\,\cdots \,+\,r_n x^n)$ by
\beq
x [N/M] = x \, \frac{1 + a_1x + ... +a_{N}x^N}{1 + b_1x + ... + b_Mx^M}~:~
x [N/M] = S + x \, {\cal O}(x^{N+M+1})
\label{two}
\eeq
i.e. the PA's are constructed so that their Taylor expansion
up to and including order $N{+}M=n$ is identical to the original series
\footnote{We use this notation, with one power of $x$ out of the brackets 
so that $S(x)$ would fit the usual notation of effective charge. Note that
the diagonal case in this notation is $x[N/N+1]$.}.
PA's may be used either to predict the next term in some
perturbative series, called a Pad\'e Approximant prediction (PAP),
or to estimate the sum of the entire series, called Pad\'e
Summation. The reasons for the success
of PA's in these different applications have not always been apparent. 
Indeed, rational functions are very flexible, and hence they are good 
candidates to approximate other unknown functions,
but this does not seem as a sufficient explanation. 
In this paper we give for the first time an argument,
based on renormalization group analysis, why 
PA's are especially well suited for summation of series describing to 
observables in QFT.

Among the areas in which PA's have had remarkable successes has been
perturbative QCD \cite{SEK,PBB}
where PA's applied to low-order perturbative series have
been shown to `postdict' accurately known higher-order terms, and also used to
make estimates of even higher-order unknown terms that agree with independent
predictions \cite{KatStr} 
based on the Principle of Minimal Sensitivity (PMS) \cite{PMS}
and Effective Charge (ECH) \cite {ECH} techniques.

In recent papers \cite{PBB,BjRS} we used the Bjorken sum rule for 
deep-inelastic scattering of polarized electrons on polarized nucleons as  
a testing ground and showcase for the use of PA's. 
We showed that applying the appropriate PA to the Bjorken Sum Rule, 
the renormalization scale (RS) and scheme dependence are significantly
reduced.
This observation provided 
circumstantial evidence that PA are resumming correctly the most 
important part of higher-order perturbative corrections. However, in
absence of an explicit mathematical argument, we could not show
that this success was not coincidental. 

In this paper we provide the missing mathematical argument and prove
that in the large $\beta_0$ limit 
(cf. Eq. (\ref{beta_func}) and (\ref{large_b0}))
when only 1-loop renormalization
of the coupling is important (e.g. large-$N_f$ QCD)
diagonal PA are {\em exactly} RS invariant, giving the   
{\em same} result, regardless of the RS. This results directly from
the fact that in the large $\beta_0$ limit the RS transformation of
the coupling reduces to a homographic transformation of the Pad\'e
argument. Diagonal PA's are invariant under such transformations. 
Non-diagonal PA's are not 
totally invariant, but we show that they reduce the RS
dependence significantly.

In general the $\beta$ function includes higher-order perturbative 
corrections which alter the running of the coupling-constant with
respect to the 1-loop evolution.
Because of this, PA's are
not exactly invariant under the RS transformation. However, since   
in QCD with $N_f=3,\, 4$ or $5$, the 1-loop running of the coupling
is dominant, PA's are still almost RS invariant.

The physical significance of the (approximate) RS invariance of PA's
is clear: a part of the contribution of the unknown high orders in a
perturbative series is due to diagrams which renormalize the
coupling-constant. The numerical importance of these terms is reflected in the
RS dependence of the partial-sum. Therefore the fact that PA's
significantly reduce the RS dependence implies that they 
correctly re-sum these terms.   
In the large $\beta_0$ limit 
(for example, large $N_f$ QCD) the arbitrariness in setting the RS 
can be interpreted as the freedom to replace the gluon propagator by one
that is corrected by an arbitrary number of 1-loop insertions. These
include both fermion bubbles, and gluon (and ghost) loops.
The fact that diagonal 
PA's become {\em exactly RS invariant} in this limit, suggests that 
they provide an {\em optimal} resummation for this type of higher order
corrections (without actually calculating them). Indeed we empirically
found \cite{SEK} that PA's succeed in predicting 
higher order coefficients \cite{Broadhurst,BBB,LTM}
in the large $N_f$ limit.

An alternative method to resum higher-order terms that renormalize the
coupling constant is the BLM \cite{BLM} method, where
one uses the fact that the leading
terms in $\beta_0$ can be identified from the leading powers in
$N_f$. One then sets the scale of the leading-order term
in the series so that all the known higher-order terms that are
leading in $\beta_0$ are absorbed. 
The hope is that also unknown higher-order terms that are leading in
$\beta_0$ will be accounted for by this choice of scale.
Thus we see that at least in the large $N_f$ limit, 
PA's and BLM are very close. Indeed we found
that in this limit choosing the BLM scale is mathematically 
identical to using non-diagonal $x[0/n]$ PA's. 
This equivalence does not hold in the general $N_f$ case.

A detailed comparison \cite{BjRS} of PA's with other methods aimed at 
optimizing perturbative expansions by setting the scale and
scheme, such as ECH and PMS has shown that there are no algebraic
relations between these two methods and 
PA's. On the other hand, good numerical agreement was found in the
case of the Bjorken Sum Rule. In this
paper we show that this agreement is general, and related to the 
approximate RS invariance of PA's.

The paper is organized as follows: In section 2 we introduce the notation and
discuss RS dependence in the large-$\beta_0$ limit. In section 3 
we prove the exact invariance of diagonal PA's under RS transformations 
in this limit, and demonstrate it by a simple example.
In section 4 we discuss non-diagonal PA's and show that
their RS dependence is reduced as compared to the corresponding partial-sums.
Section 5 is devoted to the application of our results to the 
physical regime of QCD with $N_f\,=\,3$, 4 or 5 flavors. 
In section 6 we present our 
physical interpretation of the results and compare the PA method to
the ECH, PMS and BLM methods. Section 7 includes our conclusions.    

\section{RS dependence of observables and the large $\beta_0$ limit}

At any order of perturbation theory there remains 
a residual scale dependence of
the order of the next term in the series.
Thus, as one goes to higher orders, one expects the scale dependence 
to decrease.
Unfortunately, in practice observables can only be calculated up to some
finite (and usually low) order, and thus the theoretical predictions 
are ambiguous. In QCD observables have so far been calculated only up to
 next-to-next-to-leading order (at most). Since the coupling-constant
in experimentally interesting energies is quite large, this theoretical
difficulty becomes a serious problem in comparison of theory with experiment.
 
The residual RS dependence at a given order is an indication of the importance 
of higher order corrections, in the sense that they should 
compensate for this dependence, and therefore their contribution 
cannot be smaller than the ambiguity introduced by the choice of
scale.  The actual situation is  
more complicated, since the ambiguity in the choice of scale is not 
well-defined: how far a RS can be from the physical scale without
introducing new 
physics? In what renormalization scheme would we define it? and so
on. We will not deal here with these
well-known open questions any further, but we will make strong use of the RS
dependence arguments mentioned above.

Suppose we start with an effective charge\footnote{For simplicity we
  do not deal here with the more general case of 
$S(x)\,=\,x^p(1\,+\,r_1x\,+\,\cdots)$ for $p\neq1$. The generalization
of out results is, however, straightforward.} of an observable $S$ in 
some renormalizable QFT, calculated  up to some order $n+1$ in 
perturbation theory in some renormalization scheme and scale,   

\beq
S = x \left( 1+ r_1 x + r_2 x^2 + r_3 x^3 + \cdots +r_n x^n \right)
\label{S}
\eeq 
In general, both the coefficients $r_i$ and the coupling-constant
\footnote
{The notation we use is suitable for QCD, and in some cases 
we explicitly state QCD results, but the conclusions apply 
to other QFT's as well.} 
 $\, x= \alpha_s / \pi$
 depend on the renormalization scheme and scale. If renormalization is 
 self-consistent then this dependence cancels amongst the different terms 
in the series, so that the total change in $S$ when changing 
the scheme or the scale is of order $x \,{\cal O}( x^{n+1})$.  
  
More technically, the coupling-constant $x$ in (\ref{S}) obeys the 
Renormalization Group (RG) equation:
 \beq
\frac{dx}{dt}= \beta_0 x^2 + \beta_1 x^3 + \beta_2 x^4 + \cdots\,\equiv\,
\beta(x)
\label{beta_func}
\eeq
where the first two coefficients are the same in any renormalization scheme,
while higher coefficients, $\beta_2$, $\beta_3$, $ \, \ldots$ are
 renormalization scheme dependent. In fact, as was shown in
 \cite{PMS}, at any given order in perturbation theory different 
 renormalization schemes can be uniquely defined by these higher
 coefficients of the $\beta$ function.
  
In QCD the first three coefficients of the $\beta$ function 
are known \cite{b0,b1,beta2}. The first two are:
 \beq
 \beta_0 =  -\, \frac{1}{4}\left( \,{\displaystyle \frac {11}{3}}\,C_a -
 \frac {2}{3}\, N_f \right)
\label{beta_0}
\eeq
\beq
 \beta_1 =   \frac{1}{16} \left( -\,\frac {34}{3}\, C_a^{2} +
 2\,C_f\, N_f + \frac {10}{3}\, C_a \,N_f \right)
\label{beta_1}
\eeq
where in $SU(N_c)$ gauge theory $C_a = N_c$ and $C_f = (N_c^2-1)/(2N_c)$.

We are interested in studying the effect of changing the 
renormalization scale. Suppose that $x$ above refers to the 
coupling-constant being renormalized at the physical scale (for instance,
the momentum transfer $Q^2$ in deep inelastic scattering), while we denote
by $y$ another legitimate coupling-constant renormalized at
some arbitrary scale $\mu^2$. We denote by $t$ the scale shift that
corresponds to the transformation from $x$ to $y$:
\beq
t=\ln\left({Q^2 \over{\mu^2}}\right)
\label{t}
\eeq
Equation (\ref{beta_func}) can be integrated order by 
order to give the required scale-shift transformation:
\beq
x = y+ \beta_0 t y^2 +  (\beta_0^2 t^2 + \beta_1 t) y^3 + \left(
\beta_0^3 t^3 + \frac{5}{2} \beta_1 \beta_0 t^2 + \beta_2 t \right) y^4 +
 \cdots
\label{general_ss}
\eeq

We turn now to the large $\beta_0$ limit, defined by the condition
\beq
\beta_0 \gg \beta_i x^i
\label{large_b0}
\eeq
for any $i \geq 1$. This is the limit in which our main argument is given.
Condition (\ref{large_b0}) implies that only $\beta_0^k\,t^k$ terms in  
equation (\ref{general_ss}) are significant. 
An important example where this condition is realized is QCD with a very large
number of flavours (see \cite{LTM} and the Appendix). 
Since $\beta_0$ is relatively large also in other cases,
including in QCD with a few light flavours, this approximation is worth
studying.  We return to the validity of this approximation in QCD in
a later section.

It is important to point out that the question of scheme dependence 
does not appear in the large $\beta_0$ limit, 
as all the high order coefficients of the $\beta$ function are not 
important. Thus the only remaining ambiguity is due the
arbitrariness of the renormalization scale. Therefore, the study of  
the RS dependence is much simplified in this limit. 

When all the higher order terms in the $\beta$ function are neglected,
the scale shift transformation (\ref{general_ss})  
simplifies to a geometrical series:
\beq
x= y\left(1 + \beta_0 t y +  \beta_0^2 t^2  y^2 + \beta_0^3 t^3 y^3 +
 \cdots \right)
\label{large_beta_0_ss}
\eeq
which can be summed to all orders:
\beq
x = \frac{y}{1-\beta_0 t y}
\label{ss}
\eeq
In the following we briefly review some properties of this
scale shift transformation. 

We start by recalling the fact that the transformation
(\ref{ss}) admits the following mathematical
relation, which we refer to as additivity of scale-shifts: 
\beq
t_3 = t_1+t_2
\label{additivity_of_scales}
\eeq
where 
\begin{eqnarray}
&x \stackrel{t_1}{\longrightarrow} y \stackrel{t_2}{\longrightarrow}
 z \\\nonumber
&x \stackrel{t_3}{\longrightarrow} z \nonumber
\end{eqnarray}
This property implies 
that the set of all transformations (\ref{ss}) with different 
scale-shifts $t$ forms a group. It is easy to show the existence of 
a unit operator ($t=0$), the existence of an inverse transformation 
($t \, \longleftrightarrow \, -t$) and associativity.
As for the question of whether the set is a closed one: from 
the mathematical point of view -- it's clearly not closed, 
due the the existence of the Landau pole, where 
 the denominator of (\ref{ss}) is zero:
 $t_{Landau}=\frac{1}{\beta_0 x}.$
From the physical point of view we can assume that the Landau pole is not 
 reached and thus ignore this problem, concluding that RS
shifts are indeed a group. 

As we will soon be interested in applying transformations like (\ref{ss})
to partial sums and to PA's, it is important to note here
that the transformation (\ref{ss}) is a 
rational polynomial rather than a polynomial of finite order, and as such, 
it transforms a polynomial of finite order into a rational polynomial, but 
transform an a rational polynomial into another rational polynomial.

This is a crucial
observation which is the basis of the exact RS invariance of diagonal
PA's in the large $\beta_0$ limit, and the reduced RS dependence of 
non-diagonal PA's in this limit. It is also the reason why partial 
sums at a given (finite) order can never be RS invariant.

\section{RS invariance of diagonal PA's in the large $\beta_0$ limit} 

The statement we prove is that in the large $\beta_0$ limit, starting with a 
partial sum $S(x)$  of a given order $n+1$, as in eq.~(\ref{S}) 
\footnote{We take $n$ to be an odd number, in order to construct a
  diagonal PA.}, the diagonal $x [N/N+1]$ PA 
\footnote {We remind the reader that in $x [N/N+1]$ PA we  
refer to $x$ times a rational polynomial with numerator of
order $N$ and a denominator of order $N+1$, as defined in the Introduction.
 Therefore we call it a diagonal PA.} 
 of $S(x)$, such that $n = 2N\,+\,1$,
does not depend on the RS in which the given partial sum 
was calculated.

The following diagram will be helpful in illustrating our discussion:

\begin{centering}
\large{
\begin{tabbing}
partial-sum:\,\,\,\, \,\,\=\,\,\,$S(x)$\=\,\,\,$\stackrel{t}{\longmapsto}$
 \= \,\,\,$S_1(y,t)$
 \=$\,\,\,\stackrel{-t}{\longrightarrow}$\=$\,\,\,S_2(x,t)$
  \\\>$\,\,\,\,\downarrow$\>\>\,\,\,\,$\downarrow$ \>\>\,\,\,\, \\ 
PA: \>\,\,\,$P(x)$  \>\,\,\,$\stackrel{t}{\longrightarrow}$ 
\>\,\,\,$P_1(y,t)$
 \>\,\,\,$\stackrel{-t}{\longrightarrow}$ \>\,\,\,$P_2(x)$ \\ 
\>\,\,\,\,$\updownarrow$ \>\,\,\,\,\>\,\,\,$\updownarrow$
 \>\,\,\,\>\,\,\,\,$\updownarrow$ \\ 
Taylor of PA: \>\,\,\,$T(x)$  \>\,\,\,
$\stackrel{t}{\longrightarrow}$
 \>\,\,\,$T_1(y,t)$ \>\,\,\,$\stackrel{-t}{\longrightarrow}$
 \>\,\,\,$T_2(x)$ \\ 
\end{tabbing}
}
\end{centering}
\noindent
where $S(x)$ is the $(n+1)$-th order partial 
sum as in eq.~(\ref{S}), $P(x)$ is the $x [N/N+1]$ PA 
constructed from $S(x)$, and $T(x)$ is the 
infinite-order Taylor series of $P(x)$. The horizontal arrows 
represent the application of scale-shift transformation according to equation
(\ref{ss}), once from 
 $x$ to $y$ (with the scale-shift $t$) , and then back, from $y$ to
 $x$ ($-t$).
The first scale-shift in the first line is intentionally represented by a 
different arrow ($\longmapsto$) than the other 
scale shifts ($\longrightarrow$), meaning that the series in $y$ 
is truncated at the $(n+1)$-th order after applying the scale-shift 
transformation~\setcounter{footnote}{0}\footnote{
In order to calculate $S_1(y)$, one substitutes
 $x$ as a function of $y$ in $S(x)$
 according to (\ref{ss}), 
and  Taylor expands the result to its $(n+1)$-th order,
 neglecting all the higher order terms.}. 
This 
truncation does not apply to the other scale-shift transformations in 
the diagram, where an exact transformation according to (\ref{ss}) is meant.
In the following we discuss the elements of the above diagram in detail.

The starting point in a resummation process is an 
$(n+1)$-th order partial-sum which generically
can be either $S(x)$ or $S_1(y,t) $. The two partial sums give different
numerical results for the observable, but {\em a priori} $\,$ one is just as good as
the other. The numerical difference between $S(x)$ and $S_1(y,t)$ can
be calculated by applying the {\em full} inverse scale-shift 
 transformation ($-t$) to $S_1(y,t) $ which yields $S_2(x,t)$. The
 latter is numerically the same as $S_1(y,t)$ but differs from the 
original $S(x) $ by corrections of order $x {\cal O}(x^{n+1})$.

Having described the RS dependence of the partial-sums we now consider 
PA's. We will now show that diagonal PA's are RS invariant, i.e. 
$P_2(x)$ does not depend on $t $. We will explicitly show that the two
PA's -- $P(x)$ that is based on $S(x)$, and $P_2(x) $ which is calculated
 by an inverse scale-shift transformation of $P_1(y,t) $, i.e. based on
 $ S_1(y,t)$, are exactly equal! 
This invariance results from the fact that in the large-$\beta_0$
limit, RS transformations of the coupling amounts to a 
homographic transformation of the Pad\'e argument. Diagonal 
PA's are known to be invariant under such transformations 
(see Ref. \cite{Baker} and references therein).
The proof is as follows:
\begin{description}
\item{a) \,} $P(x)$ is calculated as the $x [N/N+1]$ PA of $S(x)$:
\beq
P(x)= x \frac{1 + a_1x + ... +a_{N}x^N}{1 + b_1x + ... + b_{N+1}x^{N+1}}
\label{P}
\eeq
Similarly, $P_1(y,t)$ is calculated as the $y [N/N+1]$ PA of $S_1(y,t)$. 
\item{b) \,} The scale-shift transformation (\ref{ss})
 is applied to $P(x)$, to give a function which we denote $P_1^*(y,t)$:
\beq
P_1^*(y,t)= \left(\frac{y}{1-\beta_0 t y}\right)
 \frac{1 + a_1\left(\frac{y}{1-\beta_0 t y}\right) +
 ... +a_{N}\left(\frac{y}{1-\beta_0 t y}\right)^N}{1 + b_1\left(\frac{y}
{1-\beta_0 t y}\right) + ... +
 b_{N+1}\left(\frac{y}{1-\beta_0 t y}\right)^{N+1}}
\label{P_1_star}
\eeq
We shall see that  $P_1^*(y,t)$ is equal to $P_1(y,t)$. 
\item{c) \,} By multiplying both the numerator and the denominator by 
$(1-\beta_0 t y)^{N+1}$ we see that $P_1^*(y,t)$ is actually a
 a rational polynomial of the type $y [N/N+1]$.  
\item{d) \,} $T_1^*(y,t)$ is the Taylor expansion of $P_1^*(y,t)$. The $n$
first coefficients of $T_1^*(y,t)$ are necessarily the same as those in 
$S_1(y,t)$. This is because $T_1^*(y,t)$ can also be viewed as the 
scale-shifted version of $T(x)$, which, by the way the $P(x)$ PA was 
constructed, has the same $n$ first coefficients as $S(x)$. On the other 
hand, the scale-shift transformation is such that the $n$ first coefficients
of $T_1^*(y,t)$ (or $S_1(y,t)$) depend only on the first $n$ coefficients
in $T(x)$ (or $S(x)$).     
\item{e) \,} Since an $x [N,N+1]$ PA is uniquely determined by the first 
$n$ terms of a power series ($n=2 N\,+\,1$), it follows that
  $P_1^*(y,t)$ equals to $P_1(y,t)$. 
\item{f) \,}  If we now apply an inverse scale-shift ($-t$) on $P_1(y,t)$, 
we get $P_2(x)$ which is equal to $P(x)$ (and therefore does not depend on 
$t$). This is due the existence of an exact inverse scale-shift
transformation, or putting it differently, due to additivity of scale-shifts, 
discussed in the previous section. \footnote{This additivity of scales
  breaks down when higher powers of $t$ are present in the
  transformation, due to non-negligible
  higher-order corrections to the $\beta$ function.}  
\end{description}

This completes the formal proof of the invariance of PA's in the
 large $\beta_0$ 
limit, as stated at the beginning of the section. We now present a 
simple example to illustrate how this symmetry works in practice. We start 
with a fourth order partial-sum 
($n=3$)
\beq
S(x)= {x}\, \left( \! \,1 + {{r}_{1}}\,{x} + {{r}_{2}}\,{x}^{2}
 + {{r}_{3}}\,{x}^{3}\, \!  \right) 
\label{Sn3}
\eeq   
and calculate $P(x)$ as a $x [1/2]$ PA:
\beq
 P(x)= \,{x}\, \frac { \left( \! \,{{r}_{2}} - {{r}_{
1}}^{2}\, \!  \right)  +  \left( \! \,2\,{{r}_{1}}\,{{r}_{2}
} - {{r}_{1}}^{3} - {{r}_{3}}\, \!  \right) \,{x}}{ \left( 
\! \, - {{r}_{2}}^{2} + {{r}_{1}}\,{{r}_{3}}\, \!  \right) \,{x}
^{2} +  \left( \! \, - {{r}_{3}} + {{r}_{1}}\,{{r}_{2}}\, \! 
 \right) \,{x} + {{r}_{2}} - {{r}_{1}}^{2}}
\label{Pn3}
\eeq
Expanding $P(x)$ back in a Taylor series we get $T(x)$: 
\beq
T(x) = {x} + {{r}_{1}}\,{x}^{2} + {{r}_{2}}\,{x}^{3} + {{r}
_{3}}\,{x}^{4} + {\displaystyle \frac {{{r}_{2}}^{3} - 2\,{{r}_{2
}}\,{{r}_{1}}\,{{r}_{3}} + {{r}_{3}}^{2}}{{{r}_{2}} - {{r}_{1}}^{
2}}}\,{x}^{5} + {\cal O}(\,{x}^{6}\,)
\label{Tn3}
\eeq
We identify the $r_4$ PAP as the coefficient of $x^5$ in
(\ref{Tn3}).

We calculate the RS dependence of the partial sum $S(x)$ as follows. First, we
obtain $S_1(y,t)$ by substituting $x=\frac{y}{1-t \beta_0 y}$ in $S(x)$,
expanding the result into a power series and truncating beyond the
$y^4$ term:
\begin{eqnarray}
\label{S_1n3}
&S_1(y,t) = {y} +  \left( \! \,{{ \beta}_{0}}\,{t} + {{r
}_{1}}\, \!  \right) \,{y}^{2} \,+&\!  \left( \! \,{{ \beta}_{0}}^{2}
\,{t}^{2} + 2\,{{r}_{1}}\,{{ \beta}_{0}}\,{t} + {{r}_{2}}\, \!
 \right) \,{y}^{3} + \\ \nonumber
 & & \mbox{} +  \left( \! \,{{ \beta}_{0}}^{3}\,{t}^{3} + {{r}_{3
}} + 3\,{{r}_{1}}\,{{ \beta}_{0}}^{2}\,{t}^{2} + 3\,{{r}_{2}}\,{{
 \beta}_{0}}\,{t}\, \!  \right) \,{y}^{4}
\end{eqnarray}
We now transform back to $x$ by substituting $y=\frac{x}{1+t \beta_0 x}$ 
in $S_1(y,t)$, and expanding to all orders in $x$, to get $S_2(x,t)$.
The resulting formula for $S_2(x,t)$ is 
\begin{eqnarray}
\label{S_2n3}
&S_2(x,t)= {x} + {{r}_{1}}\,{x}^{2} \, +&\! {{r}_{2}}\,{x}^{3} + {{r}_{3}}\,{
x}^{4} + \\ \nonumber
&&  \left( \! \, - 4\,{\beta_0}^3 {t}^{3}\,{{r}_{1}} -
{\beta_0}^4 {t}^{4} - 6\,{{r
}_{2}}\,{\beta_0}^2 {t}^{2} - 4\,{{r}_{3}}\,{\beta_0}{t}\,
 \!  \right) \,{x}^{5} +
{\cal O}(\,{x}^{6}\,)
\end{eqnarray}
The RS dependence of the partial sum is the difference $\Delta S(x,t)$
between $S_2(x,t)$ and $S(x)$:
\begin{eqnarray}
\label{DSn3}
&\Delta S(x,t) \,&\equiv \, S_2(x,t)\,-\,S(x) \\ \nonumber
&&=\, \left( \! \, - 4\,{\beta_0}^3 
{t}^{3}\,{{r}_{1}} - {\beta_0}^4 {t}^{4} - 6\,{{r
}_{2}}\,{\beta_0}^2 {t}^{2} - 4\,{{r}_{3}}\,\beta_0 {t}\,
 \!  \right) \,{x}^{5} +
{\cal O}(\,{x}^{6}\,)
\end{eqnarray}

We note that the $t$ dependence of $S_2(x,t)$ and thus also 
$\Delta S(x,t)$ can be quite large for certain values of $t$ (at least
for large $t$ it is clear that the contribution due to $\beta_0^4 t^4$
cannot be canceled by other terms).
On the other hand, the $x [1/2]$ PA is exactly RS invariant:
if we start with the
fourth-order partial sum $S_1(y,t)$ of Eq. (\ref{S_1n3}), 
construct a $y [1/2]$ PA $P_1(y,t)$ and
%\begin{eqnarray}
%P_1(y,t) =\\ \nonumber
%\frac{ 
% \, \left( \! \, - {{r}_{2}} + {{r
%}_{1}}^{2}\, \!  \right) \,{y} +  \left( \! \,{{r}_{2}}\,{{ \beta
%$}_{0}}\,{t} - 2\,{{r}_{1}}\,{{r}_{2}} - {{r}_{1}}^{2}\,{{ \beta}
%_{0}}\,{t} + {{r}_{1}}^{3} + {{r}_{3}}\, \!  \right) \,{y}^{2}\,
% \!
%}   
% {   - {{r}_{2}}
%  \mbox{} + {{r}_{1}}^{2} +  \left( \! \,{{r}_{3}} + 2\,{{r}_{
%2}}\,{{ \beta}_{0}}\,{t} - 2\,{{r}_{1}}^{2}\,{{ \beta}_{0}}\,{t}
% - {{r}_{1}}\,{{r}_{2}}\, \!  \right) \,{y} 
%  \mbox{} +  \left( \! \, - {{r}_{2}}\,{{ \beta}_{0}}^{2}\,{t}
%^{2} + {{r}_{2}}\,{{r}_{1}}\,{{ \beta}_{0}}\,{t} + {{r}_{2}}^{2}
% + {{r}_{1}}^{2}\,{{ \beta}_{0}}^{2}\,{t}^{2} - {{ \beta}_{0}}\,{
%t}\,{{r}_{3}} - {{r}_{1}}\,{{r}_{3}}\, \!  \right) \,{y}^{2}  
%}
%\end{eqnarray}
transform back to $x$, we get exactly the $P(x)$ of Eq. (\ref{Pn3}), 
as proven. 

We stress that also the PAP's are RS independent. For instance,
if we calculate the PA-improved 
$5$-th order partial sum in $y$, and transform 
it back to the coupling $x$, we will get exactly the same prediction
for $r_4$, as in (\ref{Tn3}), without any $t$-dependence.
This is true not only for the PAP of $r_4$ but for any higher-order
diagonal PAP. 

\section{The reduced RS dependence of non-diagonal PA's in the large
  $\beta_0$ limit}
 
In the previous section we have proved that diagonal
 $x [N/N+1]$ PA's, are exactly RS
invariant in the $\beta_0$ limit.
In this section we will see that non-diagonal $x [N/M]$ PA's are not exactly
RS invariant even in this limit. However, on the global level 
(i.e. for large variations of the scale $t$), 
their RS dependence is much reduced compared to partial sums. 

The RS dependence of non-diagonal PA's is illustrated in the following
diagram (cf. analogous diagram in the previous section):

\begin{centering}
\large{
\begin{tabbing}
partial-sum:\,\,\,\,\=\,\,\,$S(x)$\=\,\,\,\,\,\,\,\,\,\,\,\,\,\=\,\,\,$\stackrel{t}{\longmapsto}$
 \=\,\,\,\,\,\,\,\,\,\,\,\,\,\,\,\,\,\,\,\,\,\= \,\,\,$S_1(y,t)$
 \=$\,\,\,\stackrel{-t}{\longrightarrow}$\=$\,\,\,S_2(x,t)$
  \\
\>$\,\,\,\,\downarrow$\>\>\>\>\,\,\,\,$\downarrow$ \>\>\,\,\,\, \\ 
PA: \>\,\,\,$P(x)$  \>\,\,\,$\stackrel{t}{\longrightarrow}$ 
\>\,\,\,$P_1^*(y,t)$\>\,\,\,\,\,\,\,\,\,\,\,\,\,\,$\neq$\,\>\,\,\,$P_1(y,t)$
 \>\,\,\,  \>\, \\ 
\>\,\,\,\,$\updownarrow$ \>\,\,\,\,\>\>\>\,\,\,$\updownarrow$
 \>\,\,\,\>\,\,\,\, \\ 
Taylor of PA: \>\,\,\,$T(x)$  \>\>\>\,\,\,
 \>\,\,\,$T_1(y,t)$ \>\,\,\,$\stackrel{-t}{\longrightarrow}$
 \>\,\,\,$T_2(x,t)$ \\ 
\end{tabbing}
}
\end{centering}
To illustrate this, we turn now to the simplest example,
where we start with a third order series
($n=2$): 
\beq
S(x) = x\, \left(  \,1 + r_1 \,x + r_2\,x^2\,  \right) 
\label{Sn2}
\eeq
Calculating its $x [1/1]$ PA, we get 
\beq
 P(x) = \,{x}\, \frac {{{r}_{1}} +  \left( \! \,{
{r}_{1}}^{2} - {{r}_{2}}\, \!  \right) \,{x}}{{{r}_{1}} - {{r
}_{2}}\,{x}}
\label{P11}
\eeq
Expanding back in a Taylor series we obtain:
 \beq
T(x) = {x} + {{r}_{1}}\,{x}^{2} + {{r}_{2}}\,{x}^{3} + 
{\displaystyle \frac {{{r}_{2}}^{2}}{{{r}_{1}}}}\,{x}^{4} + {\cal 
O}(\,{x}^{5}\,).
\eeq
Applying the scale shift transformation to $S(x)$ yields:
\beq
S\left(\frac{y}{1-t \beta_0 y}\right) =  \frac {{y}\, \left( \! \,1 + 
{\displaystyle \frac {{{r}_{1}}\,{y}}{1 - {\beta_0 t}\,{y}}} + 
{\displaystyle \frac {{{r}_{2}}\,{y}^{2}}{(\,1 - {\beta_0 t}\,{y}\,)^{2}
}}\, \!  \right) }{1 - {\beta_0 t}\,{y}}
\eeq
which can be Taylor expanded in $y$, and truncated at fourth order, to give:
\beq
S_1(y,t) = {y} +  \left( \! \,{{r}_{1}} + {\beta_0 t}\, \!  \right) \,{y}^{2}
 +  \left( \! \,2\,{{r}_{1}}\,{\beta_0 t} + {{r}_{2}} + {\beta_0}^2
{t}^{2}\, \! 
 \right) \,{y}^{3}
\eeq
Constructing a $y [1/1]$ PA, we get:
\beq
 P_1(y,t) =\,y\, \frac { \left( \! \,{{r}_{1}} + {\beta_0 t}\,
 \!  \right)  +  \left( \! \,{{r}_{1}}^{2} - {{r}_{2}}\, \! 
 \right) \,{y}}{ \left( \! \, - 2\,{{r}_{1}}\,{\beta_0 t} - {{r}_{2}}
 - {\beta_0 t}^{2}\, \!  \right) \,{y} + {{r}_{1}} + {\beta_0 t}}
\label{P111y}
\eeq
On the other hand, if we transform $P(x)$ to $y$, using the exact
scale-shift transformation (\ref{ss}), we get:
\beq
P_1^*(y,t)=\,y\,\frac{r_1 \,+\, (r_1^2-r_1 \beta_0 t - r_2)y}
{r_1 \,-\, (r_2+2 r_1 \beta_0 t) y \,+\, (r_2+r_1 \beta_0 t)\beta_0 t y^2}
\eeq
We see that $P_1^*(y,t)$ is a {\em diagonal} rational
polynomial, of type $y [1/2]$, rather than an off-diagonal 
rational polynomial of the type $y [1/1]$, like $P_1(y,t)$ of
Eq. (\ref{P111y}).
Therefore  $P_1^*(y,t) \,\neq\,P_1(y,t)$, and 
the scale invariance property does not hold here.

In order to measure the RS dependence of the non-diagonal PA, we define 
$\Delta T(x,t)$, in analogy with $\Delta S(x,t)$ which was
introduced in Eq. (\ref{DSn3}) to measure the RS dependence of partial-sums.
\beq
\Delta T (x,t) \,\equiv\, T_2(x,t) \,-\,T(x),
\eeq
where as $T(x)$ is the Taylor expansion of $P(x)$ and
$T_2(x,t)$ is obtained by applying an inverse transformation
($-t$) to the PA-improved partial sum $T_1(y,t)$,
the Taylor expansion of $P_1(y,t)$ (see diagram). 

Going back to our example, we calculate $T_2(x,t)$, 
\beq
T_2(x,t)\,=\, x\left[1\,+\,r_1x \,+\, r_2x^2\,+\, 
\frac{-\beta_0^2 t^2 r_2 +r_2^2 +r_1 \beta_0 t r_2 +r_1^2 \beta_0^2
t^2} {\beta_0 t +r_1}\, x^3\,+\, {\cal O}(x^4)\right]
\eeq
As expected, up to $r_2$ we obtain the same coefficients we started
with. The predicted $r_3$ turns out to be scale dependent (and
therefore different from the one obtained from $P(x)$).
Thus for the RS dependence of the PA function $\Delta T(x,t)$ we obtain:
\beq
\Delta T(x,t) \,=\,\left(\frac{-\beta_0^2 t^2 r_2 +r_2^2 +r_1 \beta_0 t r_2
  + r_1^2 \beta_0^2 t^2} {\beta_0 t +r_1}\,-\,\frac{r_2^2}{r_1}\right)
\, x^4\,+\, {\cal O}(x^5) 
\label{deltaT11}
\eeq
This has to be compared with the RS dependence of the partial sum.
Taking the inverse ($-t$) transformation of $S_1(y,t)$ we get $S_2(x,t)$:
\beq
S_2(x,t) = {x} + {{r}_{1}}\,{x}^{2} + {{r}_{2}}\,{x}^{3} +
 \left( \! \, - {\beta_0}^3 {t}^{3} - 3\,{{r}_{1}}\,{\beta_0}^2 
{t}^{2} - 3\,{{r}_{2}}\,{\beta_0}{
t}\, \!  \right) \,{x}^{4} + {\cal O}(\,{x}^{5}\,)
\eeq
and therefore:
\beq
\Delta S(x,t) \,=\,
 \left( \! \, - {\beta_0}^3{t}^{3} - 3\,{{r}_{1}}\,{\beta_0}^2
{t}^{2} - 3\,{{r}_{2}}\,{\beta_0}{
t}\, \!  \right) \,{x}^{4} + {\cal O}(\,{x}^{5}\,)
\eeq

Comparing $\Delta T(x,t)$ with $\Delta S(x,t)$ we make the following 
observations:
\begin{description}
\item{a) \,} The asymptotic behavior of $\Delta T(x,t)$ at large
  scale variations (large $t$), is much milder than that of $\Delta
  S(x,t)$
\beq
 \Delta T(x,t)\,\sim\, (r_1^2-r_2)\beta_0 t\,x^4\,+\,{\cal O}(x^5)
\eeq
vs.
\beq
 \Delta S(x,t)\,\sim\,- \beta_0^3 t^3 x^4\,+\,{\cal O}(x^5)
\eeq
\item{b) \,} On the other hand, there is a pole in $\Delta T(x,t)$. At
  certain scales it results in very large deviations of the PA from
 the value typical at most other scales. We thus have to be careful not to
 use the $x [1/1]$ PA at these scales (for any specific case one can
 plot the  $x [1/1]$ PA as a function of the RS, and identify the
 scales that are badly influenced by the pole).
\end{description}

We now turn to a second example, where we construct a $x [0/2]$ PA from
the same partial sum (\ref{Sn2}). In this case we do not give all
the details, but only state the final result for $\Delta T(x,t)$:
\beq
\Delta T(x,t) \,=\,\left[-\beta_0 t (r_1^2 - r_2)+2 r_2 r_1
  -r_1^3\right]\,x^4
\,+\, {\cal O}(x^5)
\label{deltaT02}
\eeq
We notice that for the $x [0/2]$ PA (as for the $x [1/1]$ PA) the asymptotic 
behavior is mild, but (in contrast to the $x [1/1]$ PA) 
no poles appear in $\Delta T(x,t)$. 
Thus we conclude that 
the $x [0/2]$ PA is much less scale dependent than either the partial sum
or the $x [1/1]$ PA.

An interesting result is that the asymptotic behavior of $\Delta
T(x,t)$ for large enough $t$ in the case of the $x [0/2]$ PA is {\em the same} as
that of the $x [1/1]$ PA. This can be confirmed by comparing
the large $t$ behavior of Eq. (\ref{deltaT11}) and Eq. (\ref{deltaT02}).
The mathematical reason for this similarity is that in both cases, 
the non-diagonal PA's transform under the scale-shift transformation into 
diagonal rational polynomials of order $2$ ($x [1/2]$ PA).
Since we lack one parameter ($r_3$) in order to write a ``correct'' 
$x [1/2]$ PA to describe the observable, 
we are left with some ambiguity, having a full set of functions of the
type $x [1/2]$ with coefficients that depend on one free parameter
($r_3$) rather than one specific $x [1/2]$ PA. The remaining ambiguity
is reflected both in the freedom to choose among the $x [1/1]$ and 
$x [0/2]$ PA's, and in the
freedom to set the RS. The interpretation of this result is that 
the choice between $x [1/1]$ and $x [0/2]$ PA is equivalent to the
choice between different RS's.
We will later see how this generalizes to higher-orders.

Note that throughout the analysis we did not make any specific
assumptions about the perturbative
coefficients $r_i$ of the observable under consideration. In \cite{BjRS} we 
examined the [1/1] and [0/2] PA's considered here, 
as well as the partial sum,
for the physical example  of the Bjorken Sum Rule with $N_f=3$. 
We found that indeed the RS dependence of the $x [0/2]$ PA is
much reduced as compared to the partial sum. We also found that the
$x [1/1]$ PA has a large RS dependence in specific RS's. We now see, in
retrospective from the large $\beta_0$ limit, that these features are general.
In the next section we will show why the large $\beta_0$ assumption 
is a good approximation for QCD with 3 flavours. 
But before doing so, we want to see 
how the present conclusions for the $n=2$ case can be
generalized to higher $n$.  

For $n=3$, we start with a partial sum as in Eq. (\ref{Sn3}).
We can construct two non-diagonal PA's:
$x [2/1]$ and $x [0/3]$, in
addition to the diagonal $x [1/2]$ PA  
studied in detail in the previous 
section. We found that the partial sum has a considerable RS
dependence, as described by (\ref{DSn3}), while the $x [1/2]$ PA is
exactly RS invariant. Here we look on $\Delta T(x,t)$ for the $x [2/1]$
and $x [0/3]$ PA's. 

The asymptotic behavior of  $\Delta T(x,t)$ at large scale shifts is
the same for $x [2/1]$ and $x [0/3]$ PA's:
\beq
\Delta T(x,t)\,\sim\, (r_1^2-r_2) \beta_0^2 t^2 \, x^5\,+\, {\cal O}(x^6)
\eeq
to be compared with the asymptotic behavior of $\Delta S(x,t)$
from equation (\ref{DSn3}):
\beq
\Delta S(x,t)\,\sim\,  -\beta_0^4 t^4 \, x^5\,+\, {\cal O}(x^6)
\eeq

$\Delta T(x,t)$ for
the $x [2/1]$ PA has poles (the coefficient of $x^5$ has two poles
in $t$), whereas $\Delta T(x,t)$ for the $x [0/3]$ PA does not.
This makes the use of the $x [0/3]$ PA much safer than the use of the
$x [2/1]$ PA. However, here it is preferable to use the diagonal
$x [1/2]$ PA which is exactly invariant and therefore the other two will
not be relevant. 
 
For $n=4$ we have four non-diagonal PA's: $x [3/1]$, $x [2/2]$,
 $x [1/3]$ and $x [0/4]$.
The asymptotic behavior of  $\Delta T(x,t)$ for $x [3/1]$ and $x [0/4]$ PA's
 is the same, and the leading term is proportional to 
$\beta_0^3 t^3\, x^6$; 
the asymptotic behavior of $\Delta T(x,t)$ for $x [2/2]$
 and $x [1/3]$ PA's is again the same,
 and the leading term is proportional to $\beta_0 t \, x^6$. These
 are to be compared to the asymptotic behavior of the RS dependence of the
 partial sum  ($\Delta S(x,t)$) in which the leading term is
 $\beta_0^5 t^5 x^6$. Out of the four PA's, only $x [0/4]$ does not have
 poles in  $\Delta T(x,t)$.

Generally, starting with a partial sum of order $n+1$ and
constructing an $x [N/M]$ PA ($N+M=n$), we find the asymptotic
behavior of  
\beq
\Delta T(x,t) \,\sim\, \beta_0^d t^d \, x^{n+2}
\eeq
where $d$ is given by:
\beq
d\,=\,\vert N+1-M\vert\,=\,\vert2N+1-n\vert
\eeq
while the asymptotic behavior of $\Delta S(x,t)$ is 
\beq
\Delta S(x,t)\, \sim \, \beta_0^{n+1} t^{n+1} x^{n+2}
\eeq
$d$ actually measures the dimensionality of the ambiguity space ($=\,$ number
of unknown parameters) that exists in writing a diagonal $x [N/N+1]$ PA (for
$N \geq M$) or \hbox{$x [M-1/M]$} PA (for $N < M-1$) describing the series. 
It's clear that $d<n+1$ and therefore the global RS dependence of the
$x [N/M]$ PA is milder than that of the partial sum. On the other
hand, already at the leading term in 
$\Delta T(x,t)$ there will be $N$ poles which will cause specific
RS's to exhibit sharp scale dependence, something that we want to
avoid. Only the $x [0/n]$ PA will not have any poles in $\Delta T(x,t)$.

We conclude that in general non-diagonal PA's also have a reduced
RS dependence as compared to the partial-sum. In \cite{BjRS} we
recommended considering the RS dependence in choosing the
appropriate PA. The above discussion shows that in cases
where the large $\beta_0$ approximation is valid, for series of 
even orders in the coupling-constant (odd $n=2N+1$), 
the diagonal $x [N/N+1]$ PA's should 
be preferred, while for odd orders in the coupling (even $n=2N$), 
the next to diagonal PA's ($x [N/N]$ and $x [N-1/N+1]$)
are best from the point of view
of the global RS dependence, while the $x [0/n]$ PA is the only one in
which poles in $\Delta T(x,t)$ are guaranteed not to appear, and thus
likely to exhibit the least RS dependence for small scale variations.

\section{Applicability of the results for QCD with $N_f\,=\,3$, 4 or 5}

In this section we leave the large $\beta_0$ limit and consider the 
physical example of QCD with 3, 4 or 5 quark flavors.

As mentioned in the Introduction, we already have strong evidence
 \cite{SEK,PBB,BjRS} that PA's are useful in this case.
 Moreover we showed in
\cite{BjRS} that $x [0/2]$ PA significantly reduces the RS dependence of the
 Bjorken Sum Rule in any renormalization scheme.
Therefore we already know that the results we presented
here for the large $\beta_0$
limit are approximately true also in real-world QCD, at least for the
example of the Bjorken sum rule. In this section we will show that the 
approximate invariance of PA's under RS transformations holds for a
generic QCD observable in a wide range of renormalization schemes. 
Actually, we believe that this approximate invariance holds 
 in many QFT examples, since the only requirement is that 
the running of the relevant coupling will be dominated by the 1-loop 
contribution.  
 
 In general, the scale-shift transformation (\ref{general_ss})
cannot be written in the form (\ref{ss}), 
since higher order terms in the $\beta$
 function do not vanish. 
In fact, the problem becomes even more complicated because the generic
scale-shift
transformation is not well defined, since it can only be written as 
a divergent asymptotic series, and might not even be Borel-sumable, 
due to IR renormalons.
In addition to the RS dependence, there is also 
renormalization scheme dependence, which can be viewed as arbitrariness
in setting the higher order coefficients of the $\beta$ function.
This makes it clear that the results of RS invariance in the large $\beta_0$ 
limit cannot be formally extended to the general case.

On the other hand, we can check to what extent the 
scale-shift transformation which is exact in the large $\beta_0$
limit can serve as an 
approximation to the actual scale-shift in the general case
 (\ref{general_ss}). 
Let's consider the first two terms in (\ref{general_ss}) which deviate from 
 (\ref{ss}) -- these are the third and the fourth orders in $y$:
\beq
\begin{array}{ccccl}
\beta_0^2 t^2 && \longrightarrow &&
\beta_0^2 t^2 \,+\, \beta_1 t \\
\beta_0^3 t^3 &&\longrightarrow && 
\beta_0^3 t^3 \,+\,\frac{5}{2}\beta_1\beta_0 t^2\,+\,\beta_2 t
\label{deviat}
\end{array}
\eeq
In order to measure the effect of these higher-order terms we
define:
\beq
C_1 (t) = \frac{\beta_0^2 t^2 \,+\, \beta_1 t}{\beta_0^2 t^2}
\eeq
and 
\beq
C_2 (t) = \frac{\beta_0^3 t^3 \,+
\,\frac{5}{2}\beta_1\beta_0 t^2\,+\,\beta_2 t}{\beta_0^3 t^3}
\eeq
We see that for any (non-zero) $\beta_0 $ and large enough scale shift
 $t$, the dominance of the leading $\beta_0^k t^k$
 terms is recovered. 

Since we are interested specifically in QCD with $N_f=3$, 4 and 5, 
we can numerically estimate the
dominance of the leading $\beta_0^k t^k$ terms. For instance, for
$N_f=3$ we have in the $\MSbar$ renormalization scheme \cite{b0,b1,beta2}:
$\beta_0=-2.25$, $\beta_1=-4$ and $\beta_2=-10.06$.
Using these values we calculate $C_1(t)$ and $C_2(t)$ and look at
the actual transformation of the coupling as a function of $t$.

\begin{figure}[htb]
  \begin{center}
\mbox{\kern-0.5cm
\epsfig{file=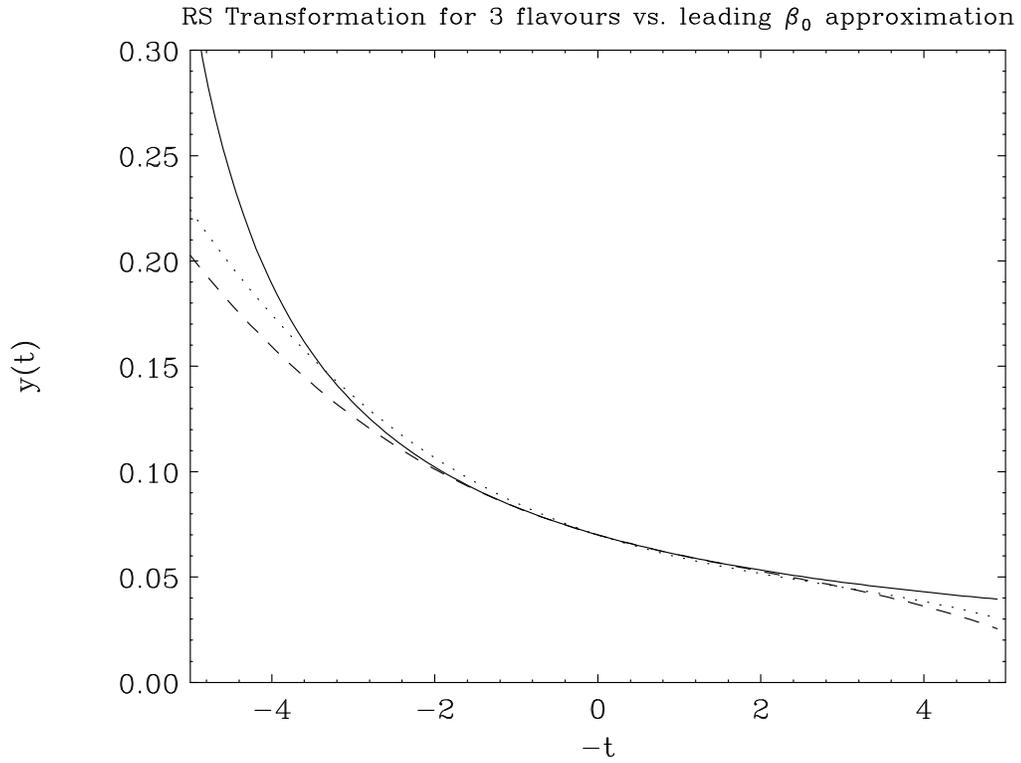,width=10.0truecm,angle=90}}
    \caption{$y(t)$ as a function of the scale-shift 
for QCD with 3 flavours, as calculated in three different ways:
(a) all-order resummation of the leading $\beta_0$ terms (continuous line),
(b) first four terms in the leading $\beta_0$ approximation (dashed line),
and (c) the first four terms in the actual QCD scale-shift transformation
(dotted line).}  
\label{Scale_shift}
  \end{center}
\end{figure}

In Figure \ref{Scale_shift} we plot $y(t)$ for 
$x \,\equiv\, \alpha_s / \pi \,=\,0.07$, corresponding 
to $Q^2\,=\,20$ GeV$^2$ in $\MSbar$ with $\mu^2 =Q^2$.
We calculate $y(t)$ in three different ways:
\begin{description} 
\item{a)\,} The exact transformation in the large $\beta_0$ limit:
\beq
y(t)\,=\,\frac {x}{1+\beta_0 t x}
\label{full_leading_bo}
\eeq
\item{b) \,} The first four terms in the leading $\beta_0$ transformation.
\beq
y(t)\,=\, x\, -\, \beta_0 t x^2\, +\, \beta_0^2 t^2 x^3\, -\, \beta_0^3 t^3 x^4
\label{4leadingb0}
\eeq
(the difference between (\ref{4leadingb0}) and (\ref{full_leading_bo}) 
at large $t$ is due to higher orders). 
\item{c)\,} The first four terms in the QCD RS
transformation, with $N_f=3$ ($\beta_2$ taken in $\MSbar$).
\beq
y(t)\,=\, x\, -\, \beta_0 t x^2\, +\, C_1(-t)\, \beta_0^2 t^2 x^3\,
 -\, C_2(-t)\, \beta_0^3 t^3 x^4
\label{QCD_trans}
\eeq
\end{description}

\begin{figure}[htb]
  \begin{center}
\mbox{\kern-0.5cm
\epsfig{file=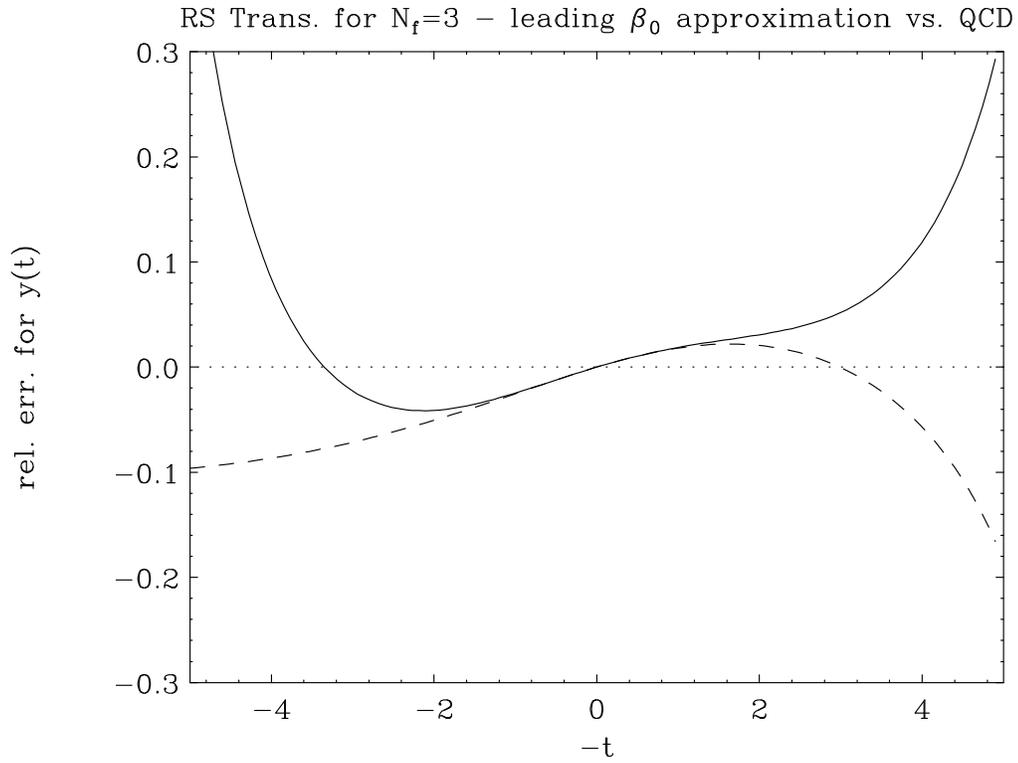,width=10.0truecm,angle=90}}
\caption{The relative deviation of the leading $\beta_0$
approximations for $y(t)$ from the QCD transformation, 
given by 
(\protect \ref{QCD_trans}).
 The continuous line represents the relative deviation of the  
all-order resummation of the leading $\beta_0$ terms
(\protect \ref{full_leading_bo}), 
while the dashed line stands for the relative deviation of the 
first four terms in the leading $\beta_0$ transformation 
(\protect \ref{4leadingb0}).
}
\label{Rel_Err_Scale_shift}
  \end{center}
\end{figure}

Figure \ref{Rel_Err_Scale_shift} presents the corresponding relative
deviations of the the leading $\beta_0$ transformations 
(\ref{full_leading_bo}) and (\ref{4leadingb0}) from the QCD 
transformation (\ref{QCD_trans}).

From figures \ref{Scale_shift} and \ref{Rel_Err_Scale_shift} 
we see that the leading $\beta_0$ approximation is quite accurate 
for QCD with 3 flavours in a very large range of scales. 
We also find that already at $\vert t \vert \gsim 4$ the error
due to the neglected higher order terms becomes
larger than the one due to neglecting of $\beta_1$
and $\beta_2$.

The numerical results for $N_f=4,5$, as well as for other
renormalization schemes (i.e. other values of $\beta_2$) 
are almost identical to those
presented in figures \ref{Scale_shift} and \ref{Rel_Err_Scale_shift}.

The conclusion from this analysis is that the scale-shift transformation 
in QCD with $N_f=3,\, 4 $ or 5 can be well approximated by the large
$\beta_0$ scale-shift transformation of (\ref{full_leading_bo}). 
Therefore, to a good approximation, the conclusions 
we drew concerning the invariance of PA's in the large $\beta_0$ limit
apply in the realistic case as well.
As mentioned, we already found through an explicit 
calculation \cite{BjRS}, that $x [0/2]$ PA reduces the RS dependence of the 
Bjorken Sum Rule in wide range of renormalization schemes.  

\section{The physical interpretation and comparison with ECH, PMS and BLM}

In the previous sections we showed that diagonal PA's become RS
invariant 
in the limit of large $\beta_0$, while non-diagonal PA's generally
reduce the RS dependence of the observable. Since the
leading $\beta_0$ approximation is good in many physical cases,
including QCD, PA's have a much reduced RS dependence there as well.
In this section we discuss the physical interpretation of this 
result and compare the PA method to the ECH, PMS and BLM methods.

\subsection{Physical interpretation}
  
In general, a significant part of the 
contribution of unknown higher orders in a
perturbative series is due to diagrams that renormalize the
coupling-constant. The numerical importance of these terms is reflected in the
RS dependence of the partial-sum since
it is the higher order terms that compensate for this unphysical dependence.

In order to understand the meaning of the (approximate) RS
independence of PA's, we first analyze it in the large $\beta_0$ limit,
where the running of the coupling-constant is completely determined by
a 1-loop renormalization. 
  
We found that in this limit diagonal  PA's become 
{\em exactly RS invariant}. 
This strongly implies that the diagonal PA sums-up the higher order
contributions that compensate for the RS dependence of the
corresponding partial-sum. 

On the other hand, even in the large $\beta_0$ limit, 
we do not expect the diagonal PA to sum  
the higher order terms {\em exactly}. This is because rational functions cannot
represent factorial behavior of the coefficients 
as expected at large orders due to renormalons.

All order perturbative calculations for some 
observables in QED and QCD in the large $N_f$ limit have been 
performed (see for instance \cite{Broadhurst} and \cite{BBB}). 
The conditions for the dominance of the 1-loop running of the
coupling-constant (\ref{large_b0}) hold in this limit (see the
Appendix), and therefore
our conclusions concerning the RS independence of diagonal PA's hold 
as well. Thus this limit provides a good testing ground for our method. 
In the large $N_f$ limit only fermion-bubble renormalon chains
contribute at high orders, and therefore the results are
extremely simple in the Borel plane. The only remaining ambiguity is
related to the integration through
poles located on the positive real axis in the Borel plane, i.e.  
the IR renormalons in QCD (and UV renormalons in QED).  After
performing the integration and going back to the $\alpha_s$ plane, we
do not get a rational polynomial in $\alpha_s$. Therefore
PA's in the $\alpha_s$ plane cannot be expected to give the ``exact''
result, but only to provide good approximations, and
``converge'' with increasing order\footnote{We use quotation
  marks to indicate that the terms are not well defined mathematically
  (due to the existence of singularities on the integration path).}.

Since the RS independence of PA's in the large $\beta_0$ limit is exact, we
suspect that diagonal PA's, although do not sum the full series,
do provide some kind of  optimal resummation of higher order corrections.
We expect, for instance, that in the large $\beta_0$ limit diagonal PA's 
are more accurate than any other non-diagonal PA and at least as
accurate as any
scale-setting method. We show later that in the large $N_f$ limit,
choosing the BLM scale is equivalent to using the $x [0/n]$
non-diagonal PA.

Given the RS independence of PA's, it is interesting to examine numerically 
the convergence of increasing order PA's and the precision of the 
Pad\'e Approximant predictions (PAP's) as 
compared to exact calculations in the large-$N_f$ limit. 
An empirical study of this kind has been done in \cite{SEK} and
\cite{PBB}. It was found that increasing order PA's do not converge
but {\em oscillate} around the Cauchy Principal-Value of the inverse Borel 
integral.  These oscillations are due to IR renormalons and in the
absence of such, increasing order PA's {\em converge} to the Borel sum of
the asymptotic series. It was also found that increasing order PAP's become 
very close to the exact perturbative coefficients,
and that the errors decrease exponentially with order. 
The errors can be approximated by a simple function,
and thus a further increase in the PAP precision is possible 
(this issue has been discussed in \cite{PBB} and in \cite{superpade}).

We stress that the exact RS independence of diagonal PA's (and the reduced
dependence  of non-diagonal PA's)  holds whenever the 1-loop running of
the coupling is dominant (cf. Eq. (\ref{large_b0})). 
Large $\beta_0$ {\em does not} necessarily imply large $N_f$, 
as there may be many other cases in QFT where the condition (\ref{large_b0})
holds. As we saw in the previous
section, the large $\beta_0$  approximation is quite good in 
QCD with only 3 to 5 flavours.  However, in contrast with QED, the
1-loop running of the coupling in QCD is dominated by gluon (and
ghost) loops and not by fermion loops.
The BLM \cite{BLM} method resums 1-loop insertions that renormalize
the coupling-constant. Therefore it is natural to compare the PA and
BLM methods, as we do in Sec. [6.3].

We see that the basic physical idea behind PA's,
and the scale and scheme setting
 methods is the same: part of the contribution of
unknown higher order corrections is related to the running of the
coupling constant. Thus it may be possible improve the perturbative
result through resumming part of these
unknown terms by relying on the characteristics of 
the renormalization group. In the following subsections we compare the
PA method to ECH, PMS and BLM.

\subsection{Comparison of PA's with ECH and PMS}

The method of Effective Charges (ECH) \cite{ECH} is based
on choosing the renormalization scale and scheme, such that the series
reduces to a leading order term (all the other known coefficients in
this scheme are exactly zero). There is a unique scheme that fulfills
this criterion.

In \cite{BjRS} we found good numerical 
agreement between $x [0/2]$ PA of the
Bjorken sum-rule  (which turned out to be almost RS invariant), and the
ECH (and PMS) scheme-setting methods. As shown in  \cite{BjRS},
there seems to be no algebraic relation between 
these two methods and the PA.
Still, our results in this paper imply that the numerical agreement is general.
The reason is quite simple, and it is directly related 
to the approximate RS independence of the PA's.
Suppose a PA was exactly scale and scheme invariant. Then
we would get exactly the same numerical result in any scheme and
scale, and in the ECH in particular. However, in the ECH scheme,
the series reduces to a leading order term
(all the higher order coefficients are exactly zero) and therefore 
the PA is identical to the ``partial sum''. Therefore
the ECH result is exactly equal to PA in any scale and scheme. Of
course, this exact agreement breaks down as effects of RS
dependence are turned on; these include the possible choice of a 
non-diagonal PA, higher order corrections to the $\beta$ function, and
scheme dependence. Still, we expect that at scales and schemes not too
far from ECH, the PA result will be much closer to the ECH result
than to the corresponding partial-sums
\footnote{This is true as long as pole effects of non-diagonal PA's
  are avoided.}. 

Another approach to set the scale and scheme is the 
Principle of Minimal Sensitivity \cite{PMS}. In
this method, one chooses the scheme in which the
renormalization scheme and scale dependence vanishes exactly. 
PMS is close to ECH both in its nature,
and in it's numerical predictions (see \cite{KatStr} and
\cite{BjRS}). Knowing now that ECH and PA's methods generally agree, we
also expect PMS to be close to PA's. 
We note that reducing the RS
dependence is the common basis for both PA's and PMS. The difference is
that PA's can be applied at any scale and scheme and they reduce the RS
dependence {\em globally} (even for large scale variations), while PMS
chooses a scale and scheme such that the {\em local} scale dependence vanishes.

\subsection{Comparison between PA's and BLM} 

The BLM method \cite{BLM} 
is based on the observation that in the large-$N_f$ limit of
QCD all the higher-order corrections are due to fermion loops, which
are also responsible for the running of the coupling.
Therefore they can be absorbed by changing the RS at
which the coupling is defined. It is reasonable to expect 
that absorbing these corrections in the scale will improve the
perturbative result.

In QCD with only a few flavours, one can use the fact that 
$\beta_0$ is linear in $N_f$ to single out
the 1-loop corrections to the coupling by identifying 
the terms that are leading in $N_f$. Thus also in QCD 
1-loop corrections to the coupling can be absorbed 
by setting the scale of the leading
term. Technically this is done by setting the scale 
so that at higher orders all the leading terms in $N_f$ 
will cancel exactly. 

BLM was generalized to account for non-leading corrections to the
renormalization of the coupling 
in several ways (see, for instance, \cite{CSR}, \cite{GruKat}
and \cite{Rathsman}).  Since the basic intuition of BLM and its
generalizations relies on the large $N_f$ limit, 
we present a detailed comparison with PA's for this
case only.  

We start with an effective charge of a generic observable in the large
$N_f$ limit (see the Appendix). 
The BLM scale-setting procedure is based on eliminating the
$N_f$ dependence of the coefficients $r_i$. In the large
$N_f$ limit this results in {\em complete elimination} of the $r_i$'s,
since in this limit $r_i \propto {N_f}^i$ (there are no
sub-leading terms), leading to the result:
\beq
S_{BLM}\, =\, x(t_{BLM}) 
\label{x_BLM}
\eeq
or, equivalently, using the large $N_f$ limit notation
in the Appendix:
\beq
S_{BLM}^{N_f \rightarrow \infty}\, =\, z(t_{BLM}) 
\label{z_BLM}
\eeq
where $z \,\equiv\, x N_f$ and 
\beq
t_{BLM}\,=\, t_1\,+\,t_2 z\,+\, t_3 z^2\,+\,\cdots\,+\,t_n z^{n-1}
\label{t_BLM}
\eeq 
where $t_1$ is proportional to $r_1$, $\, t_2\,=\,t_2(r_1,r_2)$, 
$\, t_3\,=\,t_3(r_1,r_2,r_3)$, etc.

The leading-order BLM scale $t_1$ is chosen such that $c_1$ (or
 $r_1\equiv c_1N_f$) is eliminated. 
Using this scale  
results \cite{BLM} in a summation of the 
leading diagrams which correct the gluon propagator at higher orders.
For instance, a contribution like  $c_1^2\,z^3$ (or $r_1^2\,x^3$)
is accounted for (cf. (\ref{ss_largeNf_tur})),
although terms of order $z^3$ were not initially included. 

In order to eliminate also the next coefficient $c_2$, one has to
alter the scale-shift by adding a term $t_2 z$ that is proportional to the
coupling\footnote{We will later discuss the effects of the non-leading
corrections to $t_{BLM}$.}. 
Similarly, one adds 
$t_3 z^2$ to eliminate $c_3$, and so on.
In such a way all the {\em known} terms can
be absorbed into the definition of the coupling-constant, hopefully summing
correctly the bulk of higher-order unknown contributions. 

Several different proposals were made for generalizing BLM beyond the 
leading scale $t_{BLM}\,=\,t_1$, but  in the large $N_f$ limit 
they all agree: both the single-scale BLM
  method of Ref. \cite{GruKat} and the multi-scale method of
  Ref. \cite{CSR} reduce then to (\ref{z_BLM}) and
 (\ref{t_BLM}) \footnote{This
   consensus does not include the method described in \cite{Rathsman}
   where $t_{BLM}\,=\,t_1$ at any order. In this method
   the higher order terms in $(\beta_0 x)$ as well as effects due to 
   sub-leading running
   of the coupling are resummed by setting the higher orders of
   the $\beta$ function, i.e. choosing the renormalization scheme such
   that the remaining coefficients reach their conformal-limit value. 
   This idea cannot be applied in the large $N_f$ limit,
   where one neglects the higher order corrections to the $\beta$ function
   altogether.}.

Suppose we want to calculate the effective charge in the large
$\beta_0$ limit by the BLM prescription, according to (\ref{z_BLM}) and
(\ref{t_BLM}), assuming the coupling-constant at the
physical scale (i.e. $z$) is known. The effective charge we are
calculating is just the 
coupling-constant at the scale $t_{BLM}$, given by the inverse of relation 
(\ref{ss_largeNf}):
\beq
S_{BLM}^{N_f \rightarrow \infty} (z)\,=\, 
\frac{z}{1\,+ \, \beta_0 t_{BLM} z}
\label{S_BLM}
\eeq
where from now on $\beta_0$  stands for to
$\beta_0^{N_f \rightarrow \infty}\,=\,\frac{1}{6}$. 
We substitute (\ref{t_BLM}) into (\ref{S_BLM}) to get:
\beq
S_{BLM}^{N_f \rightarrow \infty} (z)\,=\, 
 \frac{z}{1\,+\,\beta_0 t_1 z\,+\,\beta_0 t_2 z^2 \,
+\,\cdots \,+\,\beta_0 t_n z^n}
\label{S_BLM_PA}
\eeq
 
By construction, if $S_{BLM}^{N_f \rightarrow \infty}(z)$
 is expanded in powers of $z$ up to order $n+1$, 
one would get the original series $S(z)$.
We note that the r.h.s. of (\ref{S_BLM_PA}) is a $z [0/n]$ rational
polynomial. Now, since there is a unique $z [0/n]$ PA
which has an $(n+1)$-th order Taylor expansion equal to $S(z)$, we conclude
that {\em in the large $N_f$ limit BLM is  exactly equivalent to $z [0/n]$ PA's}. 

In view of this result it is worthwhile to repeat the
characteristics of the $z [0/n]$ functions, found in Section 4:
\begin{description}
\item{a) \,} As other non-diagonal PA's,  a $z [0/n]$ PA
{\em does} depend on the RS. The leading term in this RS
 dependence is proportional to $\beta_0^{n-1} \, t^{n-1}$,
 significantly less than $\beta_0^{n+1} \, t^{n+1}$ in the partial-sum.
\item{b) \,}  As opposed to other non-diagonal PA's, the $z [0/n]$ PA
  is strictly {\em free} from any poles in its RS dependence.
\end{description}

We wish to emphasize that the BLM prescription beyond the leading
order (i.e. beyond $t_{BLM}\,=\,t_1$) cannot be regarded as a
choice of RS in the strict sense,
 since terms that depend on $z$ in (\ref{t_BLM})
break the additivity of scale-shifts in the transformation
(\ref{ss_largeNf}). Putting it differently: after substituting $t_{BLM}$ in
(\ref{ss_largeNf}), $w$ depends on $z$ in a more
complex way than implied by the RS transformation. 
This is why the BLM result is not free of RS dependence and is not
equivalent to a {\em diagonal} PA~\footnote{The argument we used in
  subsection 6.2 to show
the agreement between ECH and PA's does not apply in the case
of BLM in the large $N_f$ limit, even though the series reduces to a
single term. The reason is precisely the fact that
 BLM is not strictly a choice of RS.}.
This does not apply to the simplest case of a next-to-leading order
series, where $t_{BLM}=t_1$. Here, BLM in the large $N_f$ limit becomes simply
 a $x [0/1]$ PA, which is RS invariant. 

We wish to stress that the equivalence of $x [0/n]$ PA's and BLM is of
course true {\em only} in the large $N_f$
limit. Moreover, $x [0/n]$ PA's are different from all the
various generalization of BLM that attempt to take into account
effects due to non-leading running of the coupling-constant. If one
wishes, one can regard $x[0/n]$ PA's as another generalization of this kind.

\section{Conclusions}

We showed that in the $\beta_0$ limit diagonal PA's of perturbative 
series become {\em exactly} renormalization scale independent.
 
This implies that diagonal PA's are correctly resumming  
contributions from higher order diagrams which are responsible for the 
renormalization of the coupling-constant.
   
Non-diagonal PA's are not exactly invariant even in the large
$\beta_0$ limit, but still reduce the global RS dependence as
compared to partial-sums. Among the different non-diagonal PA's, the
only one that has a completely  regular behavior with respect to 
scale variations is 
the $x [0/n]$ PA. We have shown that in the large $N_f$ limit of QCD, 
the latter is identical to the BLM scale-setting procedure.

In physical cases, higher order corrections in the $\beta$ function break
the RS independence of PA's, introducing a small scale and scheme
dependence, even for diagonal PA's.

We also showed that PA's, when they are indeed RS invariant to a good
approximation, lead to the same numerical result as the ECH method. 

An important feature of PA's (one that in our view makes them more useful 
than scale and scheme setting methods) is simply that they 
can be used in any scale and scheme. 
The comparison of results in different scales and schemes
can then serve two goals:
\begin{description}
\item{a) \,} Estimate the reliability of the PA method is in each
  particular case, by 
 considering the scale and scheme dependence of the partial sum as a 
 reference, as was done for the Bjorken sum rule in Ref. \cite{BjRS}. 
\item{b) \,} Use the residual scale and scheme dependence as a lower
  bound for the theoretical error.
\end{description}

Finally, we feel that the ``surprising success'' of PA's in QCD, 
and generally in QFT, is now based on a much more firm basis.

\bigskip
\begin{flushleft}
{\bf Acknowledgements}
\end{flushleft}
I thank Marek Karliner and John Ellis for very useful discussions.
The research 
was supported in part by the Israel Science Foundation
administered by the Israel Academy of Sciences and Humanities,
and by
a Grant from the G.I.F., the
German-Israeli Foundation for Scientific Research and
Development.

\vskip 14pt
\newpage
\setcounter{equation}{0}
\def\theequation{A.\arabic{equation}}
{\Large
\bf{Appendix - The large $N_f$ limit}
}

In this Appendix we briefly present some of the basic formulae that
are used in the large $N_f$ limit calculations. This concerns the
discussion in section 6 and especially subsection 6.3 where we compare
PA's and BLM in this limit.

Starting with (\ref{S}) and using the fact that the leading term in
$r_i$ is proportional to ${N_f}^i$, we obtain:
\begin{eqnarray}
& S(x) &= x \left( 1+ r_1 x + r_2 x^2 + r_3 x^3 + \cdots +r_n x^n
\right)\\ \nonumber
&&= x \left( 1+ c_1 N_f x + c_2 N_f^2 x^2 + c_3 N_f^3 x^3 + \cdots +c_n N_f^n
   x^n \right)
\label{S_largeNf}
\end{eqnarray}
We define $z \equiv x N_f$ and
 $S^{N_f \rightarrow \infty}(z)\,=\,S(x)N_f$ and thus: 
\beq
S^{N_f \rightarrow \infty}(z)\,=\, 
z \left( 1+ c_1 z + c_2 z^2 + c_3 z^3 + \cdots +c_n z^n \right)\\
\label{Sz}
\eeq

In the same manner we change the notations for the 
$\beta$ function (\ref{beta_func}):
\beq
\frac{dz}{dt}= \frac{\beta_0}{N_f} z^2 + \frac{\beta_1}{N_f^2} z^3 + 
\frac{\beta_2}{N_f^3} z^4 + \cdots
\label{beta_func_largeNf}
\eeq

Remembering that $\beta_0\, \sim \, N_f$ while at higher orders 
$\beta_i \sim {N_f}^i$
(cf. (\ref{beta_0}) and (\ref{beta_1})),
 we conclude that
higher order corrections to the $\beta$ function are negligible, being
sub-leading in $N_f$, and (\ref{beta_func_largeNf}) can be written as:
\beq
\frac{dz}{dt}= \beta_0^{N_f \rightarrow \infty} z^2 
\label{beta_func_z}
\eeq
where $\beta_0^{N_f \rightarrow \infty}\,=\,\frac{1}{6} $.
Therefore Eq. (\ref{large_beta_0_ss}) translates into:
\beq
z\,=\, w \,+\,\beta_0^{N_f \rightarrow \infty} t  w^2 
\,+\,\left(\beta_0^{N_f \rightarrow \infty}\right)^2 t^2  w^3 
\,+\,\left(\beta_0^{N_f \rightarrow \infty}\right)^3 t^3  w^4 \,+\,\cdots
\label{ss_largeNf_tur}
\eeq 
where $w$ is the coupling-constant that is defined at the
 new renormalization point $\mu$ (see Section 2).
Finally, Eq. (\ref{ss}) translates into:
\beq
z\,=\,\frac{w}{1\,-\,\beta_0^{N_f \rightarrow \infty} t  w }
\label{ss_largeNf}
\eeq 

\newpage

\vskip 14pt

\def\etal{{\em et al.}}
\def\PL{{\em Phys. Lett.\ }}
\def\NP{{\em Nucl. Phys.\ }}
\def\PR{{\em Phys. Rev.\ }}
\def\PRL{{\em Phys. Rev. Lett.\ }}

\newpage

\newpage

\end{document}